\documentclass[letterpaper]{article}

\usepackage[T1]{fontenc}

\usepackage{geometry}
\usepackage{setspace}
\usepackage[unicode=true]{hyperref}
\hypersetup{breaklinks=true, colorlinks=true, urlcolor= blue, citecolor=blue}
\usepackage{nameref}
\usepackage{amsmath}

\usepackage[style = chem-acs, doi=true]{biblatex}
\usepackage{graphicx}
\usepackage{float}
\newfloat{scheme}{htbp}{los}
\floatname{scheme}{Scheme}
\floatname{chart}{Chart}
\newfloat{graph}{htbp}{loh}

\usepackage{chemformula} 
\usepackage[version = 4]{mhchem} 

\usepackage{authblk}
\author[1]{Vijay Yadav}
\author[1]{Sunita Kumari*}
\author[1]{Prabhat K. Jaiswal*}
\affil[1]{Department of Physics, Indian Institute of Technology Jodhpur, Jodhpur, Rajasthan 342 030, India}

\title{Machine Learning Accelerated Simulation of Charge Regulation Effects}
\date{*Email: prabhat.jaiswal@iitj.ac.in, sunita@iitj.ac.in}

\begin{document}

\maketitle

\begin{abstract}
Many solvated materials, from nanoparticles to polyelectrolytes, acquire charge through the dissociation or association of surface groups, a phenomenon known as charge regulation (CR). Recent methods use hybrid charge regulation Monte Carlo (CR-MC)/molecular dynamics (MD) to evolve the system while dynamically adjusting charges on the surface groups. However, this approach is computationally expensive, as Monte Carlo sampling can be as costly as the accompanying MD. In this work, we introduce charge regulation machine learning (CR-ML), a graph attention network that predicts the ionization probability of every surface site from the instantaneous configuration, the dissociation constant of surface groups, and the solution conditions. We show that CR-ML reproduces the equilibrium conformations and degree of ionization obtained with CR-MC for a hydrophobic polyanion. We further test CR-ML on the more complex system of two nanoparticles mediated by a polyelectrolyte, where it accurately predicts the charge asymmetry that develops between the two particles. Across these systems, CR-ML updates charges an order of magnitude faster than CR-MC while accurately reproducing its equilibrium properties. This framework opens a path toward large-scale simulations of charge-regulating materials for understanding and designing their responsive behavior.

\end{abstract}

\section{\label{sec:into}Introduction}

Coulomb interaction rules our lives. It serves as an architectural force behind myriads of biochemical processes such as enzyme catalysis, cellular and nucleic acid organization, gene regulation, and nerve signal transmission~\cite {bian21npassembly, muthu23qmacromols}. Interestingly, many biomolecules and their synthetic counterparts are often decorated with dissociable charged groups~\cite{muthu23qmacromols}. Depending on surrounding physicochemical signals, these charged moieties dynamically alter their fate and charge state. This fascinating charge adjustment is called charge regulation (CR)~\cite{muthu23qmacromols, grant2001, elcock2001, lund2005, lund2013,amin2019, curk2021, curk22, yuan2022, kumari24screencrmion, Yuan2024, kandari26screencrzwit, beyer2026, Yadav2026, Tyagi2026-ps, Kaur2026-xu, You2026, Xiong2025}. CR plays a paramount role in understanding protein interactions~\cite{grant2001, elcock2001, lund2005, lund2013}, colloidal stability~\cite{amin2019, vangara2026, szilagyi2012}, nanoparticles and viral capsid assembly ~\cite{Yadav2026, Rik14, Ivan24, LpezFlores2025},  to name a few. The multi-stimuli responsive sensitivity of CR has revolutionized various fields from modern therapeutics to smart material design~\cite{Han2020, Jiang2011, Ritt2022}. Despite the importance of CR in a plethora of applications, most studies still employ the traditional constant charge (CC) approximation, as it is relatively simple and entails lower computational costs. However, several studies have shown that outside the conventional CC scenario, CR can yield intriguing results. For example, anomalous ion transport~\cite{Rangharajan2025}, charge symmetry breaking~\cite{Majee2018,Tyagi2026-ps, Itani2026}, and notable binding affinity by 30-fold~\cite{beyer2025}. This striking difference highlights that incorporating CR is essential for accurately predicting the behavior and properties of weakly ionizable systems.

The theoretical treatment of CR dates back nearly a century, beginning with Linderstrøm-Lang's work on ionization equilibria~\cite{linderstromlang1924}. Building on this foundation, Kirkwood and Shumaker showed that fluctuations in the ionization states of nearby groups can generate electrostatic attractions beyond those predicted for fixed charges~\cite{kirkwood1952}. Subsequent developments relied on the Poisson–Boltzmann description of the electrolyte and were largely limited to relatively simple geometries~\cite{ninham1971, Podgornik2018}. Particle-based simulations offer more accurate representation and great versatility but at substantially higher computational cost, as modeling acid–base reactions requires ab initio molecular dynamics (AIMD) to compute interatomic forces on the fly~\cite{Lu2008, Maurer2010}. Consequently, such simulations are limited to very short time spans and small length scales. 

Constant pH Monte Carlo (MC)~\cite{Reed1992, Radak2017, Izzo2025} and reaction ensemble MC (RxMC)~\cite{Smith1994, Johnson1994, Turner2008} are two commonly used coarse-grained techniques to model acid-base equilibria. The constant-pH method treats pH as a fixed input and neglects dissociated protons and hydroxide ions explicitly, so it is valid only when the salt concentration far exceeds their concentration, whereas RxMC models these ions explicitly but can only exchange them through matching reservoir salt species, causing severe finite-size errors near neutral pH. To bridge this gap, various methods have been proposed, such as the Grand-reaction (G-RxMC)~\cite{Landsgesell2020} and charge regulation Monte Carlo (CR-MC)~\cite{curk22}. Despite these advances, the computational cost of such methods remains a major bottleneck for large systems. For example, in CR-MC, the total cost of MC sampling scales as $O(MN\log N)$ with the number of ionizable sites $M$ and charges $N$, overtaking the cost of the molecular dynamics (MD) propagation itself in systems with many dissociable groups, thereby limiting the accessible system sizes and timescales even with this accelerated hybrid scheme.

Recent developments in machine learning (ML) offer methods to accelerate computer simulations~\cite{behler2007, botu2015, behler2016, chmiela2018, zhang2018, bigi2025, liu2017, huang2017, nagai2017, shen2018, chen2018}, thereby enabling access to timescales and system sizes previously unreachable. ML has been used to construct fast interatomic potentials for long-time MD~\cite{behler2007, zhang2018}, accelerate sampling of rare and slow conformational transitions~\cite{bonati2019, noe2019}, extend coarse-grained models to larger length and time scales~\cite{wang2019, zhang2018deepcg}, and learn efficient move proposals that speed convergence in MC simulations~\cite{liu2017, huang2017}. In particular, graph neural networks (GNNs) have proven well suited to disordered, many-body particle systems, as they naturally encode the instantaneous local environment of each particle while respecting permutation and translational symmetry~\cite{scarselli2009, kipf2017, gilmer2017, battaglia2016, battaglia2018, sanchezgonzalez2020, bapst2020}. Such architectures have been used to predict per-atom energies and forces in machine-learned interatomic potentials~\cite{schutt2017, batzner2022} and to predict local physicochemical properties, such as partial charges, directly from molecular structure~\cite{unke2019}, at a small fraction of the cost of the underlying physical calculation. In the context of CR, GNNs have been trained to predict protonation states of titratable sites in proteins and small molecules from a fixed structure~\cite{cai2021, johnston2023}, showing that ionization state is learnable
directly from the structure.

In this work, we introduce charge regulation machine learning (CR-ML), a graph attention network that replaces the iterative Monte Carlo sampling of CR-MC with a single forward pass. The model takes the instantaneous particle configuration, represented as a graph, together with the intrinsic dissociation constants of the surface groups and the solution conditions, and directly predicts the ionization probability of every surface site. This prediction is used to update charges within the same hybrid MD workflow as CR-MC, without requiring the hundreds of sequential MC moves needed to equilibrate the charge distribution at each configuration. We first validate CR-ML on a weak polyanion system, in which it reproduces chain conformations across a range of solution conditions. We then apply it to a more complex system in which both cationic nanoparticles (NPs) and an anionic polyelectrolyte (PE) undergo CR simultaneously, and show that CR-ML captures cooperative CR effects, such as symmetry breaking between the NPs and the PE. CR-ML reduces the cost of charge sampling by an order of magnitude compared to CR-MC and brings the overall simulation cost close to that of a standard MD simulation. To our knowledge, this is the first ML approach that efficiently samples ionization states on the fly during MD, allowing charges to respond continuously to the evolving configuration. This opens the way to large-scale simulations of charge-regulating systems, such as protein solutions, PE complexes, and colloidal mixtures.

\section{\label{sec:methods}Methods}
\subsection{\label{sec:cr-mc}CR-MC Method}
This section outlines the CR-MC method of Curk and Luijten, which serves as the reference throughout this work~\cite{curk22}. The system is an implicit-solvent electrolyte at constant temperature $T$ containing acidic groups $A$, basic groups $B$, and free monovalent ions, in equilibrium with a reservoir of fixed pH and salinity. The groups acquire charge through the ionization reactions $A\rightleftharpoons A^-+\mathrm{H}^+$ and $B\rightleftharpoons B^++\mathrm{OH}^-$, and the reservoir additionally exchanges the products of water self-dissociation and salt ion pairs with the system. Instead of treating each of these species separately, all monovalent cations are grouped into a single type $X^+=\{\mathrm{H}^+,\mathrm{S}^+\}$ and all monovalent anions into $X^-=\{\mathrm{OH}^-,\mathrm{S}^-\}$. Provided the ions within a group share the same interaction potential, the grouping leaves the grand-canonical distribution of charge states unchanged and reduces the set of reactions to three,
\begin{align}
  A &\rightleftharpoons A^- + X^+, \label{eq:rxn_acid}\\
  B &\rightleftharpoons B^+ + X^-, \label{eq:rxn_base}\\
  \emptyset &\rightleftharpoons X^+ + X^-, \label{eq:rxn_ion}
\end{align}
which markedly improves the sampling efficiency near neutral pH.

The reservoir enters only through the chemical potentials of the exchanged species. These are written in the usual base-10 logarithmic form, so that the protons are specified by $\mathrm{pH}=-\beta\mu_{\mathrm{H}}\log_{10}e$ and the salt ions by $\mathrm{p}I_{S^\pm}=-\beta\mu_{S^\pm}\log_{10}e$, with $\beta=1/k_BT$. Here $\mu_{\mathrm{H}}$ and $\mu_{S^\pm}$ are the chemical potentials of the protons and of the salt ions in the reservoir. A smaller $\mathrm{p}I_{S^\pm}$ corresponds to a more concentrated reservoir. The intrinsic dissociation constants follow in the same way from the dissociation free energies of a single group, $\mathrm{p}K_a=\beta\Delta G_a\log_{10}e$ and $\mathrm{p}K_b=\beta\Delta G_b\log_{10}e$, where $\Delta G_a$ and $\Delta G_b$ are the free energy changes for dissociating one acidic and one basic group. They fix the ionization that a group would adopt if it felt no other charges, so that $\mathrm{pH}-\mathrm{p}K_a$ measures how strongly the reservoir drives an isolated acid to dissociate. Each of the three reactions may be attempted in either direction, and the six resulting moves are accepted by a Metropolis criterion that weighs this chemical driving force against the change $\Delta E$ in the potential energy of the system.

The method is used as a hybrid scheme with molecular dynamics: the configuration is propagated for $n_\mathrm{MD}$ steps at fixed charges, $n_\mathrm{MC}$ moves are then attempted at fixed positions, and the two alternate for the length of the run. Because the ionization state is resampled at the instantaneous configuration, every group responds through $\Delta E$ to the local electrostatic environment created by all other charges, which is the charge regulation effect that a constant charge (CC) model cannot capture. A state point is therefore specified by the pH of the reservoir, the intrinsic constants $\mathrm{p}K_a$ and $\mathrm{p}K_b$ of the groups, and the reservoir salinity $\mathrm{p}I_{S^\pm}$. We use the open-source implementation of the authors, \texttt{fix charge/regulation} in LAMMPS~\cite{LAMMPS, curk22}.

\subsection{\label{sec:cr-ml}CR-ML Method}
\begin{figure}[t]
    \centering
    \includegraphics[width=0.85\linewidth]{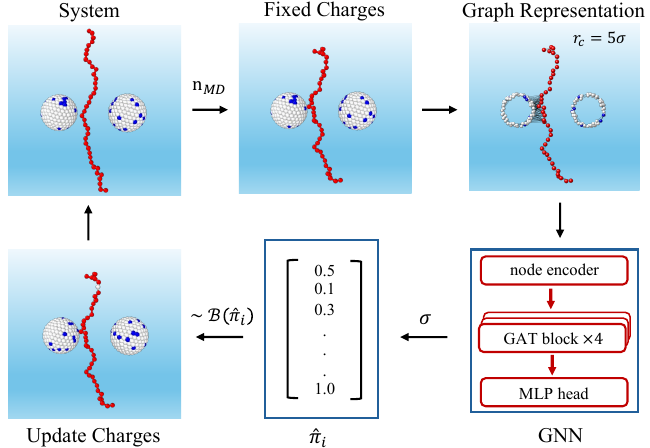}
    \caption{\label{fig:cr_ml_pipe} The workflow of charge regulation machine learning (CR-ML) scheme.}
\end{figure}
We propose a charge regulation machine learning (CR-ML) framework that replaces the hundreds of attempted Monte Carlo moves of \nameref{sec:cr-mc} with a single forward pass. Unlike the sequential Markov chain it replaces, this evaluation is highly parallel and therefore much faster on a GPU. The working of the proposed method is illustrated in \autoref{fig:cr_ml_pipe}. Starting from any system configuration, we run $n_\mathrm{MD}$ molecular dynamics (MD) steps at fixed charges, map the resulting snapshot onto a graph, and pass it through a graph neural network (GNN) that returns an ionization probability $\hat{\pi}_i$ for every site. The state of each site is drawn from a Bernoulli distribution with that probability, its charge is set accordingly, and the cycle repeats for the rest of the simulation.

Let $\mathbf{R}=\{\mathbf{r}_i\}$ be the instantaneous positions of the $N$ dissociable sites and $\mathbf{s}=\{s_i\}$, $s_i\in\{0,1\}$, their protonation state, with $s_i=1$ for an ionized site. At fixed configuration, the CR-MC ensemble defines a conditional distribution $P(\mathbf{s}\,|\,\mathbf{R};\mathbf{c})$, with $\mathbf{c}$ collecting the thermodynamic control parameters. We define the per-site marginal
\begin{equation}
\pi_i(\mathbf{R};\mathbf{c}) = \big\langle s_i \big\rangle_{P(\mathbf{s}|\mathbf{R};\mathbf{c})},
\label{eq:target}
\end{equation}
as the equilibrium ionization probability of site $i$. We train the GNN to predict this quantity directly from the configuration and the control parameters, and draw the charges from the factorized distribution
\begin{equation}
P(\mathbf{s}\,|\,\mathbf{R};\mathbf{c}) \simeq \prod_{i=1}^{N}\hat{\pi}_i^{\,s_i}\,(1-\hat{\pi}_i)^{1-s_i}, \qquad s_i \sim
\mathcal{B}(\hat{\pi}_i),
\label{eq:factorize}
\end{equation}
with $\hat{\pi}_i$ the network output. Equation~\eqref{eq:factorize} is the only approximation introduced beyond the accuracy of the network itself. The instantaneous correlations between the charge states of different sites at fixed $\mathbf{R}$ are discarded, while their mean and its spatial variation are retained, since $\hat{\pi}_i$ is a function of the entire neighborhood of site $i$.

The graph is constructed as follows. Its nodes are the dissociable sites, and its edges connect every pair of nodes closer than a cutoff $r_c$. It is rebuilt at every CR-ML call, so the network always sees the current contact topology. A node carries its chemical identity (acid or base), the corresponding offset of the solution pH from the intrinsic dissociation constant, the salt condition of the reservoir, and the charge it carries in the snapshot. Any remaining control parameter of the state point enters as an additional node feature, so that a single network covers the whole condition grid. An edge carries a Gaussian radial-basis expansion of the pair distance,
\begin{equation}
  e^{(k)}_{ij} = \exp\!\left[-\frac{(r_{ij}-\mu_k)^2}{2w^2}\right], \qquad \mu_k = \frac{k\,r_c}{K-1}, \quad w = \frac{r_c}{K-1},
  \label{eq:rbf}
\end{equation}
with $k=0,\dots,K-1$ and $K=4$. Only interparticle distances enter the representation, so the prediction is invariant under translation, rotation, and relabelling of the sites, and the same trained network can be applied to geometries it has not been trained on.

The node features are embedded by a shallow encoder into node states $\mathbf{h}$ of hidden width $F=96$ and passed through four graph attention blocks of the GATv2 form\cite{brody2022how}, which admit the edge features into the attention score
\begin{align}
\alpha_{ij} &= \operatorname*{softmax}_{j\in\mathcal{N}(i)} \Big[\mathbf{a}^{\top}\mathrm{LeakyReLU}\big(\mathbf{W}_1\mathbf{h}_i + \mathbf{W}_2\mathbf{h}_j + \mathbf{W}_e\mathbf{e}_{ij}\big)\Big], \label{eq:gat_alpha}\\
\mathbf{m}_i &= \sum_{j\in\mathcal{N}(i)}\alpha_{ij}\,\mathbf{W}_2\mathbf{h}_j. \label{eq:gat_msg}
\end{align}
Here $\mathcal{N}(i)$ is the set of nodes within the cutoff $r_c$ of site $i$, together with $i$ itself, and $\mathbf{e}_{ij}$ is the radial-basis edge feature of Eq.~\eqref{eq:rbf}. The matrices $\mathbf{W}_1$ and $\mathbf{W}_2$, of size $F\times F$, and $\mathbf{W}_e$, of size $F\times K$, project the receiving node, the neighboring node, and the edge onto a common space, where they are summed and passed through a LeakyReLU of negative slope $0.2$. The learned vector $\mathbf{a}$, of length $F$, contracts the result to one score per edge, and the softmax normalizes these scores over the neighborhood, so that $\sum_{j\in\mathcal{N}(i)}\alpha_{ij}=1$. Because the nonlinearity acts before the contraction with $\mathbf{a}$, the ranking of the neighbors by attention weight depends on the receiving site rather than being common to all sites, so that two sites of different chemical identity may weight the same neighbor differently. The same projection $\mathbf{W}_2$ carries the value in Eq.~\eqref{eq:gat_msg}, so the message $\mathbf{m}_i$ is the attention-weighted mean of the projected neighbor states, through which the local environment that shifts the ionization equilibrium of site $i$ enters the prediction. Equations~\eqref{eq:gat_alpha} and~\eqref{eq:gat_msg} are evaluated by four attention heads, each carrying its own $\mathbf{W}_1$, $\mathbf{W}_2$, $\mathbf{W}_e$ and $\mathbf{a}$ and each returning a message of the full width $F$, and the four messages are averaged. 

Each block adds the averaged message to the node state as a residual update, $\mathbf{h}_i\leftarrow\mathrm{LN}\big[\mathbf{h}_i+\mathrm{SiLU}(\mathbf{m}_i)\big]$ with $\mathrm{LN}$ the layer normalization, followed by a position-wise feed-forward network of inner width $2F$ applied to every node with the same weights and added as a second residual update of the same form. A two-layer perceptron maps the final node embedding to a scalar logit $z_i$, and $\hat{\pi}_i=\sigma(z_i)$ with $\sigma$ the logistic function, which constrains the output to $[0,1]$ by construction. The network has $4.7\times10^{5}$ parameters. Four rounds of message passing give each site a receptive field of $4r_c$. 


\subsection{\label{sec:data-train}Dataset Preparation and Training}
Training data are generated in two stages. In the first stage, one CR-MC/MD trajectory is run per condition, with the CR-MC activated every $n_\mathrm{MD}$ MD steps for $n_\mathrm{MC}$ CR-MC attempted moves. From each run, $200$ configurations are stored at equal intervals. Each stored configuration is then run for $n_\mathrm{relax} = 4n_\mathrm{MD}$ MD steps with charges held fixed, so that the final configuration is no longer equilibrated with its charge distribution.


In the second stage, each selected configuration is labeled with the target of Eq.~\eqref{eq:target}. To this end, we restart the simulation from the configuration saved in the first stage and perform $d_\mathrm{MC}$ CR-MC moves with the particle positions held fixed, that is, with the MD removed. The number of moves is chosen long enough for the charge state to decorrelate from the stored charges, and the final charge state is recorded. For each configuration, we perform $N_\mathrm{ens} = 100$ such independent runs, and the ionization probability of each site is given by,
\begin{equation}
\pi_i = \frac{1}{N_\mathrm{ens}}\sum_{e=1}^{N_\mathrm{ens}} s_i^{(e)}, \qquad \mathrm{SE}(\pi_i) = \sqrt{\frac{\pi_i(1-\pi_i)}{N_\mathrm{ens}}},
\label{eq:label}
\end{equation}
and the binomial standard error of that estimate is retained as a per-site measure of the ionization probability uncertainty, which sets the accuracy floor that any model can reach on these data.

The training objective combines a binary cross-entropy on the soft target with a mean squared error on the probability,
\begin{equation}
\mathcal{L} = -\frac{1}{M}\sum_i\Big[\pi_i\ln\hat{\pi}_i + (1-\pi_i)\ln(1-\hat{\pi}_i)\Big] + \frac{\lambda}{M}\sum_i(\hat{\pi}_i-\pi_i)^2,
\label{eq:loss}
\end{equation}
with $\lambda=0.25$ and the sums run over the $M$ nodes of the graph. The cross entropy is the natural objective for a Bernoulli variable and dominates the gradient far from the target, while the quadratic term controls the absolute error near $\hat{\pi}\to0$ and $\hat{\pi}\to1$, where most sites sit and to which the sampling of Eq.~\eqref{eq:factorize} is most sensitive. We optimize with AdamW (learning rate $10^{-3}$, weight decay $10^{-4}$) using an effective batch of $32$ graphs, a dropout of $0.05$ and bfloat16 mixed precision. We halve the learning rate when the validation error stops improving, and stop early on the same criterion. The checkpoint of the lowest validation error is retained and evaluated once on the test split. Further, the data are split $80/10/10$ by condition rather than by frame, so all frames from a trajectory remain in the same split. Since frames within a trajectory are strongly correlated, a frame-wise split would introduce near-duplicate configurations across splits and yield overly optimistic errors. In contrast, the grouped split provides a more rigorous assessment of generalization to state points that are not represented in the training data.

\section{Results}
\subsection{Conformations of a hydrophobic weak polyanion}
To evaluate CR-ML, we first apply it to the hydrophobic weak polyanion studied by Curk et al.~\cite{curk22}, in which the equilibrium conformation of a single chain is set by the competition between hydrophobic attraction and the electrostatic repulsion of its own ionized groups. The chain is a sequence of $N_m=80$ weak acid groups, each a sphere of diameter $\sigma$ and mass $m$, in an implicit solvent whose Bjerrum length equals the bead diameter, $\ell_B=\sigma=0.72$ nm, corresponding to water at room temperature. Time is reported in units of $\tau=\sigma\sqrt{m/k_BT}$. All non-bonded pairs interact through a shifted and truncated Lennard-Jones potential given by
\begin{equation}
U_{LJ}~(r_{ij}) =
\begin{cases}
4\epsilon_{\mathrm{LJ}} \left[ \left(\dfrac{\sigma}{r_{ij}}\right)^{12} - \left(\dfrac{\sigma}{r_{ij}}\right)^6 + C \right], & r_{ij} \leq r_{\mathrm{cut}}, \\[8pt]
0, & r_{ij} > r_{\mathrm{cut}}.
\end{cases}
\label{eq:lj}
\end{equation}
The energy $\epsilon_{\mathrm{LJ}}$ of Eq.~\eqref{eq:lj} takes a different value for each pair type. Monomer pairs use the full cutoff $r_{\mathrm{cut}} = 2.5\sigma$ with $C = 2.5^{-6} - 2.5^{-12}$, so their well depth $\epsilon_{mm}$ sets the strength of the effective hydrophobic attraction and is one of the two parameters we scan. All other pairs are purely repulsive, with $r_{\mathrm{cut}} = 2^{1/6}\sigma$ and $C = 1/4$. Free ions interact among themselves with strength $\epsilon_{ii} = k_BT$ and with the monomers with $\sqrt{\epsilon_{mm}\epsilon_{ii}}$, following the Lorentz--Berthelot mixing rule~\cite{lorentz1881,berthelot1898}. Neighboring monomers along the chain are bonded through a harmonic potential,
\begin{equation}
U_{\mathrm{bond}}(r_{ij}) = K(r_{ij}-R_0)^2,
\label{eq:bond}
\end{equation}
with spring constant $K=400k_BT/\sigma^2$ and rest length $R_0=2^{1/6}\sigma$.

\begin{figure}[h]
\centering
\includegraphics[width=0.90\linewidth]{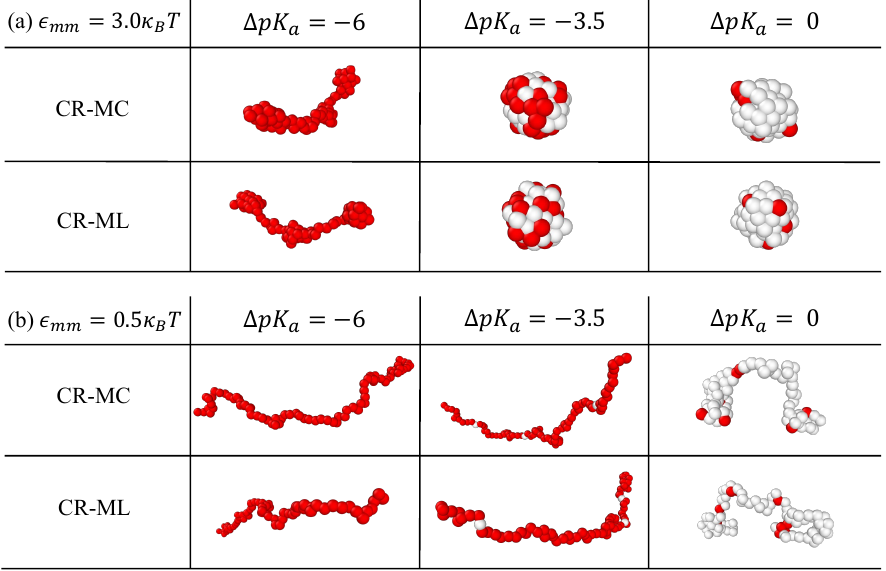}
\caption{\label{fig:pol_conf} Representative equilibrium conformations of the hydrophobic weak polyanion, from the CR-MC (upper row) and from the CR-ML (lower row), at (a) strong monomer attraction $\epsilon_{mm}=3.0\,k_BT$ and (b) weak attraction $\epsilon_{mm}=0.5\,k_BT$. Each case is shown for dissociation strengths $\Delta \mathrm{pK}_a = -6$, $-3.5$, and $0$. Ionized monomers are drawn in red and neutral monomers in white; free ions are omitted for clarity.}
\end{figure}

The chain is built as a helix of radius $3\sigma$ inside a cubic periodic box of edge $L = 100\ell_B$. The helix places every bond at $R_0$ and leaves no overlapping non-bonded pair. The box is then filled at random with the monovalent cation--anion pairs with no ion closer than $R_0$ to an existing particle. Free ions carry charge $\pm1$ and a monomer carries $0$ or $-1$ according to its protonation state. Electrostatic interactions are treated by the particle–particle particle–mesh (PPPM) solver with a relative force accuracy of $10^{-5}$ and a real-space Coulomb cutoff of $25\ell_B$. The equations of motion are integrated with the velocity-Verlet algorithm at a time step of $0.005\,\tau$, and the system is thermostatted by a Langevin bath at $k_BT=1$ with a damping time of $1\,\tau$. Further, the reservoir is held at $\mathrm{pH}=7$, $\mathrm{pI}_s=3$, and the chemical potentials of the exchanged ions follow from the neutrality condition of \nameref{sec:cr-mc}. 

The training data for this system is generated with the following settings, i.e., $n_{MD} = 400$, $n_{MC} = 200$, and $d_{MC} = 600$. The two parameters that are scanned are the $\epsilon_{mm}$ and $\Delta \mathrm{pK}_a = \mathrm{pK}_a - \mathrm{pH}$, which measures how strongly the reservoir drives an isolated group to dissociate. The CR-ML is trained on the $35$ conditions formed by $\epsilon_{mm}/k_BT \in \{0.5,1,2,3,5\}$ and integer $\mathrm{pK}_a$ from $1$ to $7$. Further, it is tested on a finer grid of $77$ conditions, $\epsilon_{mm}/k_BT \in \{0.5,1,1.5,2,3,4,5\}$ and $\mathrm{pK}_a$ in steps of $0.5$ between $3$ and $7$ against the CR-MC.

Over the state points we sample, the chain adopts four of the five conformations reported for this system~\cite{curk22, Ulrich2005}, namely the pearl necklace, the collapsed globule, the extended string, and the random walk, and the CR-ML reproduces each of them. Representative configurations of both methods are shown in \autoref{fig:pol_conf} at two interaction strengths and three dissociation constants.

\autoref{fig:pol_conf}a compares polyanion chain conformations from CR-ML and CR-MC at strong monomer--monomer attraction, $\epsilon_{mm}=3.0\,k_BT$. At strong dissociation, $\Delta\mathrm{p}K_a=-6$, the chain is nearly fully ionized. The competition between electrostatic repulsion and hydrophobic attraction produces a pearl-necklace structure, with compact beads connected by short, stretched segments. At intermediate dissociation, $\Delta\mathrm{p}K_a=-3.5$, partial ionization weakens the repulsion, allowing the chain to collapse into a compact but still appreciably charged globule. At weak dissociation, $\Delta\mathrm{p}K_a=0$, most groups are neutral, resulting in a dense and nearly uncharged globule. \autoref{fig:pol_conf}b shows the corresponding comparison at weak monomer--monomer attraction, $\epsilon_{mm}=0.5\,k_BT$. Here, the chain remains open across all dissociation strengths. At $\Delta\mathrm{p}K_a=-6$ and $-3.5$, it is nearly fully ionized and adopts an extended-string conformation. At $\Delta\mathrm{p}K_a=0$, it becomes an essentially neutral random walk.

\begin{figure}[t]
    \centering
    \includegraphics[width=0.85\linewidth]{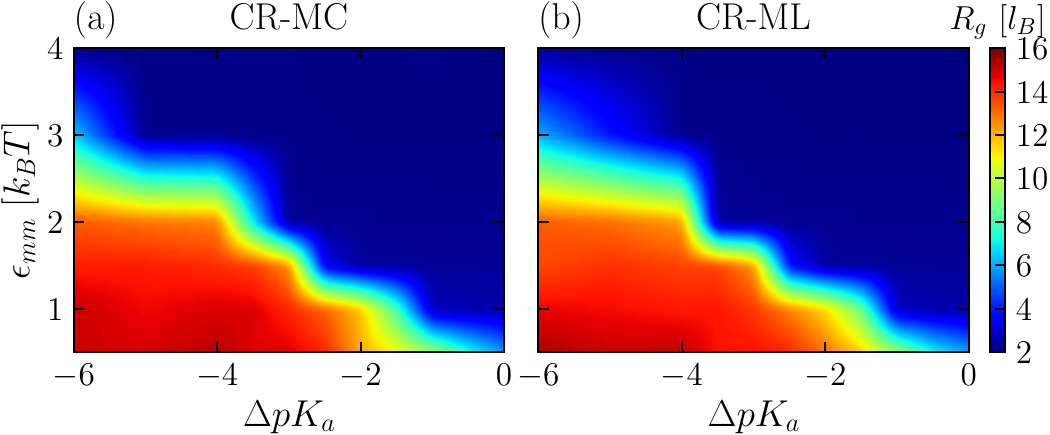}
    \caption{\label{fig:rg} Radius of gyration, $R_g$, of the hydrophobic weak polyanion over the plane of interaction strength, $\epsilon_{mm}$, and dissociation strength, $\Delta \mathrm{pK}_a$, from (a) the CR-MC and (b) the CR-ML.}
\end{figure}

At the same $\Delta\mathrm{p}K_a=-3.5$, the weakly attracting chain is nearly fully ionized, whereas the strongly attracting chain is only partially ionized. Collapse brings the ionized groups closer together, increasing the cost of further deprotonation and creating a feedback between conformation and ionization. The CR-ML reproduces this asymmetry, including in unseen conditions, because it predicts the ionization of each configuration rather than imposing a prescribed charge state. Since the two methods generate independent trajectories, the agreement in \autoref{fig:pol_conf} should be interpreted at the level of the conformational class.

Next, we present the analysis of radius of gyration, $R_g$, across the parameter space spanned by $\epsilon_{mm}$ and $\Delta\mathrm{pK}_a$ in \autoref{fig:rg}. The CR-MC result is shown in \autoref{fig:rg}(a), and the corresponding CR-ML prediction in \autoref{fig:rg}(b). Both maps exhibit an extended regime at weak attraction and strong dissociation, where the chain expands to $R_g\simeq15\,\ell_B$ (red shading), and a collapsed regime at strong attraction or weak dissociation, where $R_g$ saturates at $2.2\,\ell_B$ (blue shading). The transition between these regimes is abrupt, with $R_g$ decreasing by a factor of approximately seven within a single grid spacing. The CR-ML reproduces the location of this crossover across the parameter space. Across the 77 conditions, the mean absolute difference between the two maps is $0.22\,\ell_B$, corresponding to approximately 4\% of the local $R_g$, and only eight conditions show a difference exceeding $0.5\,\ell_B$. All of these points lie near the crossover, where $R_g$ varies steeply with the ionized fraction, such that small errors in $\alpha$ can produce comparatively large deviations in the chain conformation.

\begin{figure}[t]
    \centering
    \includegraphics[width=0.85\linewidth]{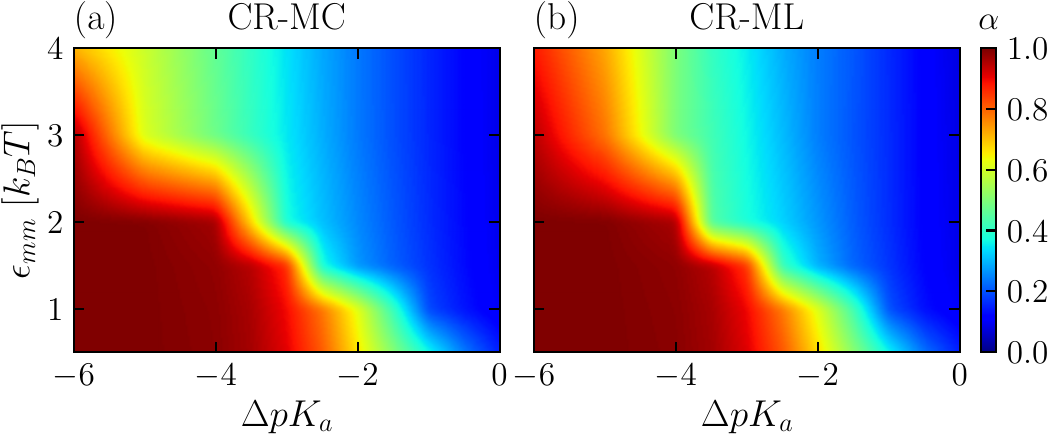}
    \caption{\label{fig:alpha} Degree of dissociation, $\alpha$, the fraction of monomers carrying $q=-1$, over the same plane and with the same construction as \autoref{fig:rg}, from (a) the CR-MC and (b) the CR-ML.}
\end{figure}

We next examine the degree of dissociation $\alpha$ over the same parameter space, from the CR-MC in \autoref{fig:alpha}(a) and from the CR-ML in \autoref{fig:alpha}(b). In both, $\alpha$ falls smoothly as $\Delta \mathrm{pK}_a$ increases. The open chain is fully dissociated at $\Delta \mathrm{pK}_a=-6$ (red shading) and is left barely charged at $\Delta \mathrm{pK}_a=0$, where $\alpha\simeq 0.1$ (blue shading). The CR-ML follows this variation closely over the whole plane. The two maps differ by $0.018$ in $\alpha$ on average, and only six of the $77$ conditions differ by more than $0.05$.

Finally, we analyze the speedup gained by replacing the CR-MC steps with the CR-ML surrogate. We time the two charge updates on the same set of configurations, on a single CPU core and with the network evaluated on that same core, so that no GPU is assumed. One CR-MC update of $n_\mathrm{MC}=200$ moves takes about $430$ ms. One CR-ML update takes $7$ ms for an open chain and $90$ ms for the densest globule, the difference coming from the number of edges in the graph, which grows as the chain collapses. Per update, the CR-ML is therefore $5$ to $60$ times cheaper than the CR-MC. It is important to note that this speedup applies only to the charge update; the cost of the MD propagation is unchanged.

\subsection{Polyelectrolyte adsorption on nanoparticles}
Having established the CR-ML on a single chain, we now apply it to a system in which two charge-regulating cationic nanoparticles (NPs) are mediated by one weak anionic polyelectrolyte (PE), following the setup of Yadav et al~\cite{Yadav2026}. We choose this system because both objects are charge-regulated, and each does so in the other's field. A PE that adsorbs on an NP lowers the electrostatic cost of ionizing the basic sites underneath it, so the particle it sits on becomes more charged than the one it has left, and the charge of every object in the box depends on where the PE happens to be. The CR-ML must therefore predict two chemically distinct site species simultaneously, on a graph far denser than that of a single PE.

Each NP is coated with 256 base groups characterized by a dissociation constant $pK_b$ and has an outer-shell radius of $4\ell_B$. The two NPs are held fixed in a cubic periodic box with edge length $80\sigma$ by excluding them from the MD integration steps. The PE consists of 40 weak acid monomers, each carrying a charge of either 0 or $-1$. The monomers are bonded according to eq~\eqref{eq:bond}, with $K=100k_BT/\sigma^2$ and $R_0=2^{1/6}\sigma$. Every non-bonded pair interacts through the purely repulsive form of Eq.~\eqref{eq:lj}, with $\epsilon_{\mathrm{LJ}}=k_BT$, $r_{\mathrm{cut}}=2^{1/6}\sigma$ and $C=1/4$, so that this system carries no hydrophobic attraction and the only attraction in it is electrostatic. Coulomb interactions are treated by the PPPM solver with a relative force accuracy of $10^{-3}$ and a real-space cutoff of $12.5\sigma$, and the equations of motion are integrated with the same time step and Langevin bath as in the previous system.

The reservoir is held at $\mathrm{pH}=7$, and all three reactions are now active: the acid reaction on the monomers, the base reaction on the NP surface sites, and the salt reaction. The CR-ML is trained on the $192$ conditions formed by $\mathrm{pK}_a$ and $\mathrm{pK}_b$ in $\{1,3,5,7\}$, $\mathrm{pI}_s$ in $\{3,5,7\}$ and $D/\ell_B$ in $\{3,7,11,15\}$, from which $1.2\times10^{4}$ labelled configurations are drawn. The training data is generated following the method described in section \nameref{sec:data-train} with the following settings, i.e., $n_{MD} = 400$, $n_{MC} = 200$, and $d_{MC} = 2000$. Further, the trained CR-ML model is compared against the CR-MC on a separate grid of separations, $D/\ell_B=2,4,\ldots,20$, at $\mathrm{pK}_a=\mathrm{pK}_b=5$ and at two ionic strengths, $\mathrm{pI}_s=3$ and $\mathrm{pI}_s=6$, with 20 independent replicates of each condition. None of these separations appears in the training grid, and neither does the ionic strength $\mathrm{pI}_s=6$, so every point of the comparison asks the network to extrapolate.

\begin{figure}[h]
    \centering
    \includegraphics[width=0.85\linewidth]{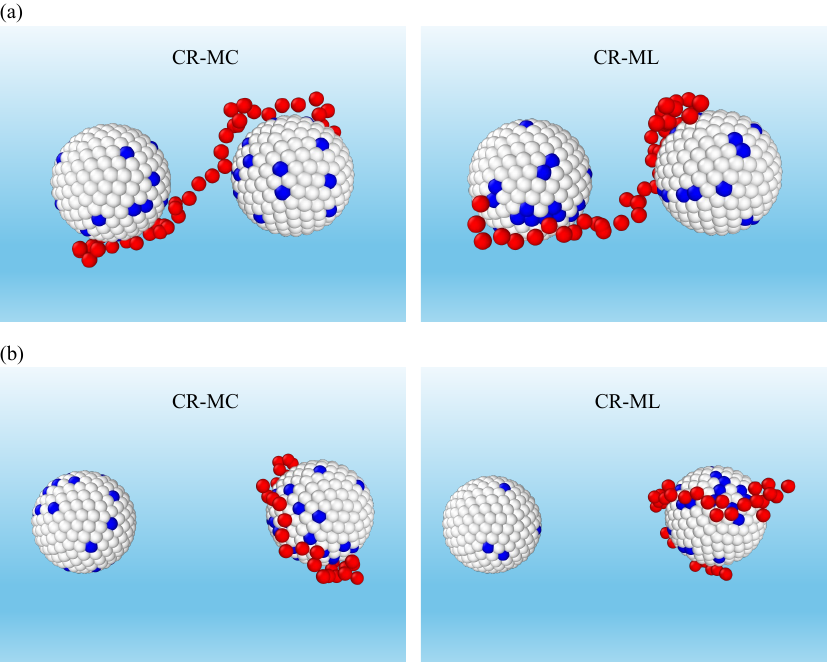}
    \caption{\label{fig:np_morph} Equilibrium morphologies of polyelectrolyte adsorption on nanoparticles at (a) intersurface separation $D = 4\ell_B$ and (b) $D = 10\ell_B$, with $\mathrm{pK}_a = \mathrm{pK}_b = 5$ and $\mathrm{pI}_s = 6$. In each case, the left panel shows the result from CR-MC, while the right panel shows the corresponding CR-ML predictions.}
\end{figure}

The graph nodes comprise the two NP cores, 512 NP surface sites, and 40 monomers, with edges connecting nodes separated by less than the cutoff distance $r_c=5\sigma$. Besides the features listed in \nameref{sec:cr-ml}, each node carries an indicator of whether it is an acid or a base site, so that the offsets $\mathrm{pH}-\mathrm{pK}_a$ and $\mathrm{pH}-\mathrm{pK}_b$ act on the monomers and on the NP surface sites, respectively. Each production run covers $2\times10^{6}$ steps, or $10^{4}\tau$, with $n_{MC} = 200$ and $n_{MD} = 400$ for CR-MC/MD, and the CR-ML updating the charges on the same interval.

To demonstrate the performance of CR-ML for the NP--PE system, we first examine the resulting morphologies in \autoref{fig:np_morph}. In \autoref{fig:np_morph}(a), the left panel shows a configuration from \nameref{sec:cr-mc}, where the PE forms a bridge between the two NPs, while the right panel shows the corresponding CR-ML configuration, which reproduces the same bridged morphology. Both configurations have an intersurface separation of $D=4\ell_B$, with $\mathrm{pK}_a=\mathrm{pK}_b=5$ and $\mathrm{pI}_s=6$. At the larger separation $D=10\ell_B$, shown in \autoref{fig:np_morph}(b), the PE instead adsorbs onto one NP in both simulations, leaving the other NP bare. All other parameters are unchanged. The two methods are run as independent trajectories, so the agreement in \autoref{fig:np_morph} is at the level of the morphology rather than frame by frame.

\begin{figure}[h]
    \centering
    \includegraphics[width=0.85\linewidth]{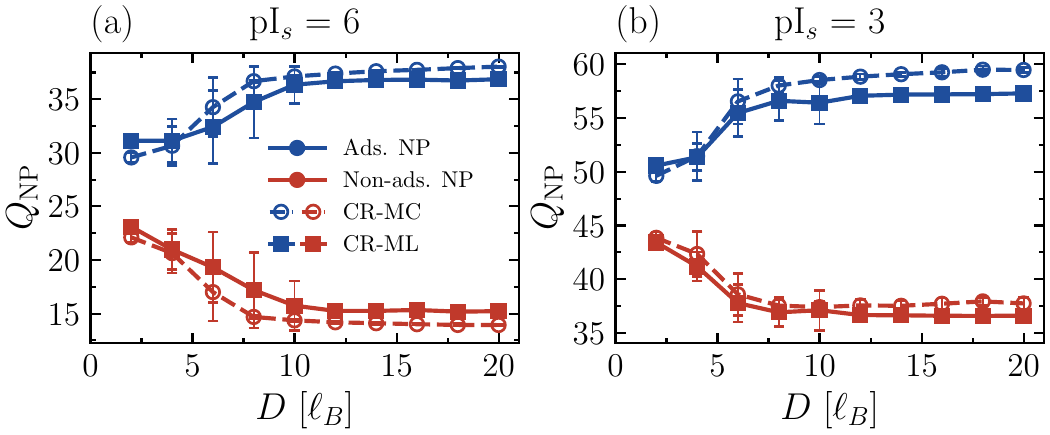}
    \caption{\label{fig:np_charges} Charge, $Q_{\mathrm{NP}}$, on each nanoparticle (NP) against intersurface separation $D$ at (a) $\mathrm{pI}_s = 6$ and (b) $\mathrm{pI}_s = 3$, with $\mathrm{pK}_a = \mathrm{pK}_b = 5$. Blue curves correspond to NP carrying the adsorbed polyelectrolyte, while red curves represent bare NP. Dashed lines with open circles are for the CR-MC method, and solid lines with filled squares are for the CR-ML surrogate. Error bars are the standard deviation over 20 independent seeds, averaged over the last 400 frames.}
\end{figure}

Next, we analyze the charge on the nanoparticles, $Q_{\mathrm{NP}}$, taken as the number of ionized base sites on each NP surface. \autoref{fig:np_charges}(a) compares CR-ML with the CR-MC at $\mathrm{pI}_s = 6$, a reservoir salt concentration of $1\,\mu$M. In both methods, the NP that adsorbs the PE acquires more charge than the bare one, and the difference widens with separation. At $D=2\ell_B$, the two carry about $30$ and $22$ charges out of the $256$ sites available on each surface, a difference of $8$. By $D=20\ell_B$, the adsorbed particle has risen to $38$, while the bare one has fallen to $14$, a difference of $24$. The sum of the two is almost unchanged across the whole range, so this is a redistribution of charge between the particles rather than a change in how much charge the pair draws from the reservoir. The split saturates beyond $D\approx12\ell_B$, which is where the chain stops bridging and commits to one NP. The CR-ML curves closely track the CR-MC across the full separation range.

\autoref{fig:np_charges}(b) presents the results at higher salt, $\mathrm{pI}_s = 3$, which corresponds to a reservoir salt concentration of $1\,$mM. The added salt screens the electrostatic interactions between the surface sites, increasing the ionization of both NPs. At the smallest separation, $D = 2\ell_B$, the two carry similar charges, about 50 on the NP with more adsorbed monomers and 43 on the other. As the separation increases, the PE adsorbs preferentially on one NP and eventually commits to it, leaving the other bare. The charges follow the same behavior in both methods, the adsorbed NP acquiring 59 ionized sites against just 38 on the bare one. At both salt concentrations, CR-ML closely matches CR-MC, capturing the charge asymmetry that develops with separation and differing from the reference by at most $2.5$ elementary charges out of the $256$ sites available on each NP surface.

\begin{figure}[h]
    \centering
    \includegraphics[width=0.85\linewidth]{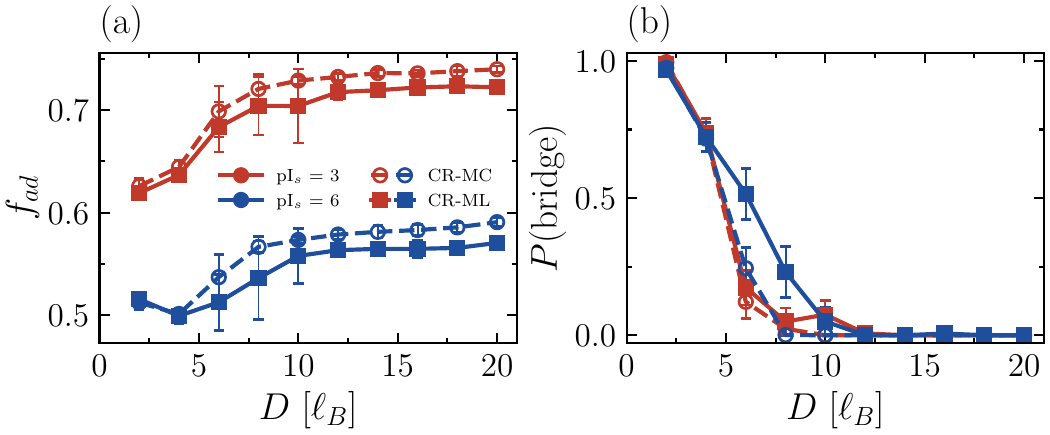}
    \caption{\label{fig:np_adsorb}(a) Adsorbed fraction $f_{ad}$ of the polyelectrolyte on nanoparticles and (b) bridging probability $P(\mathrm{bridge})$ against intersurface separation $D$, with $\mathrm{pK}_a = \mathrm{pK}_b = 5$. Blue curves correspond to $\mathrm{pI}_s = 6$, while red curves correspond to $\mathrm{pI}_s = 3$. Dashed lines with open circles denote the CR-MC method, whereas solid lines with filled squares represent the CR-ML surrogate. Error bars are the standard deviation over 20 independent seeds in (a) and the standard error of the mean in (b), averaged over the last 400 frames.}
\end{figure}

Further, we examine the adsorption of the polyelectrolyte on the NPs and the probability that it bridges them. A monomer counts as adsorbed when it lies within $1\sigma$ from the NP surface, and a configuration counts as bridged when both NPs carry at least one adsorbed monomer. \autoref{fig:np_adsorb}(a) shows the adsorbed fraction $f_{ad}$ against separation at the two salt concentrations. In both methods, $f_{ad}$ rises from $0.63$ at $D = 2\ell_B$ to about $0.73$ beyond $D = 12\ell_B$ at $\mathrm{pI}_s = 3$, and from $0.51$ to about $0.58$ at $\mathrm{pI}_s = 6$, so the whole curve lies lower at the lower salt concentration. \autoref{fig:np_adsorb}(b) shows the probability that the PE chain bridges the two NPs. This probability is nearly unity at $D=2\ell_B$, decreases sharply around $D\approx6\ell_B$, and approaches zero by $D\approx12\ell_B$ at both salt concentrations. The CR-ML closely reproduces the adsorbed fraction, running about $0.02$ below the CR-MC at every separation, and it reproduces both limits of the bridging curve. Between them, at $D=6$ to $10\ell_B$, it holds the chain bridged more often than the CR-MC does, most visibly at $\mathrm{pI}_s=6$ and $D=6\ell_B$, where it gives $0.51$ against $0.25$. These are the separations at which a bridge is metastable within a single trajectory, so that a small difference in the predicted ionization changes how long it survives.

Finally, we compare the computational cost of the two methods for this system. A single charge update using CR-MC requires approximately $0.8$ s, corresponding to the time required for $200$ MC moves at this system size. In contrast, the CR-ML performs the same update in approximately $22$ ms on a single GPU, yielding a speedup of approximately $36\times$. As before, this speedup applies only to the charge update, while the cost of the MD propagation is unchanged.

\section{Conclusions}
In this work, we propose charge regulation machine learning (CR-ML), a graph attention network that predicts the ionization state of a system from its configuration and parameters such as pH, salt concentration, and dissociation constants. CR-ML is integrated with molecular dynamics (MD) in a hybrid simulation framework, where MD updates the particle positions, and the neural network updates the charge states. We benchmark CR-ML against the charge regulation Monte Carlo (CR-MC) to assess its accuracy and computational efficiency.  We validate CR-ML on a hydrophobic weak polyanion and a system of two cationic nanoparticles (NPs) mediated by an anionic polyelectrolyte (PE). When coupled with molecular dynamics, CR-ML reproduces the CR-MC results for both systems, including at conditions outside the training data. The charge update step is $5$ to $60$ times faster for the hydrophobic polyanion system on a single CPU core and $36$ times faster for the NP--PE system on a single GPU. This reduction in the computational cost of charge regulation simulations enables more efficient studies of electrostatic interactions across soft and biological matter, including drug delivery, colloidal stability, protein interactions, and nanoparticle assembly. A key direction for future development is a network that transfers across particle geometries, molecular architectures, and chemistries without requiring retraining for each system. Viewed from a broader perspective, CR-ML provides a framework that can assist in predicting the structural and functional properties of proteins, as well as the characteristics of stimulus-triggered smart materials, a capability that could have potential applications in drug delivery and beyond.

\section*{Acknowledgments}
SK acknowledges the financial support received from the Anusandhan National Research Foundation, India (EEQ/2023/000676) and IIT Jodhpur for a research initiation grant (I/RIG/SNT/20240068). P.K.J. acknowledges the financial support from Anusandhan National Research Foundation (ANRF), India (Grant No. CRG/2022/006365) and IIT Jodhpur for a Seed Grant (No. I/SEED/PKJ/20220016).

\printbibliography




  
  

\end{document}